# Overcoming Challenges in Bamboo Connections: A Review of Mechanical Properties and Structural Considerations

Pierre Boucher[1][0009-0002-4569-0810] , Victor Fréchard[2][0000-0002-1085-1373], Diego Ramirez-Cardona[1][0000-002-1550-8048] and Claudiane Ouellet-Plamondon[1][0000-0003-3795-4791]

[1] École de technologie supérieure, Université du Québec, Montréal, Canada
[2] URM MAP, Nancy School of Architecture, Nancy, France
`pierre.boucher@woodandbamboo.fr`

**Abstract.** Over the past decades, bamboo has increasingly gained attention as a sustainable construction material, through its rapid growth, naturally optimized shape, high mechanical properties, and significant environmental benefits. However, despite these advantages, the use of bamboo in its natural form for structural applications remains limited, partly due to insufficient knowledge of connection behavior, which is crucial for ensuring the long-term reliability and performance of bamboo structures. This article provides a comprehensive review of the key factors to consider in the design of structural bamboo connections and discusses the existing connection classification methods used as guidelines by designers. By synthesizing findings from the literature, our research aims to identify the key parameters interacting with the connection design process, focusing on the anatomical, geometric, and mechanical properties of bamboo, the mechanical requirements of the structure design, and the building methods. A critical analysis of Janssen's classification of bamboo connections, based on force transfer modes and later refined by Widyowijatnoko, is presented. Finally, we discuss the identified research gaps and emphasize the need for integrated design approaches supported by guidelines to support the broader adoption of bamboo in construction.

## 1 Introduction

Bamboo is a fast-growing plant, with certain species offering mechanical properties well-suited for structural use. Naturally found in Asia and Latin America, it has been deeply embedded in the construction heritage of these regions [1]. Traditionally employed in an empirical manner for building small-scale structures and individual homes, bamboo has recently seen a resurgence of interest due to its rapid growth rate [2], its mechanical properties, its ability to meet urgent growing housing needs in regions where it is endogenous [3], its favorable environmental footprint, and its capacity for carbon storage. This renewed focus has inspired architects, engineers, and constructors to explore new tectonics and develop an architectural language with its own vocabulary, exhibiting the bamboo's potential use in construction to design and build multi-story projects or more complex structures such as gridshells.

While specialized craftsmanship is essential to the successful construction of such projects, empirical methods alone are no longer sufficient to ensure their structural performance. The integration of structural engineering has therefore become indispensable for the design and the dimensioning of bamboo structures. To date, several standards and construction guidelines provide calculation rules for the design of bamboo structures [4,5]. However, few of them offer specific calculation rules for the design of connections. Recent research has contributed to increasing the understanding of the connection's behavior [6-11] but further studies are needed to guide engineers in designing safe structures while maintaining the economic, social, and environmental viability of projects.

To address this need for support in the engineering design process, Janssen [12] proposed a classification of connections based on load transfer mechanisms, which was further refined by Widyowijatnoko [9]. While this classification provides designers with a general understanding of the underlying mechanisms, it offers limited guidance for decision-making during the early design stages of a project, failing to integrate all the factors involved in the design process.

In order to complete the previously proposed classifications, our research aims to identify the key factors to consider in the design of structural bamboo connections and then to clarify the lack in the Widyowijatnoko classification method, which will enable us to propose an improved one in further work. The paper first reviews the key factors to consider in the design of structural bamboo connections, essentially based on material properties and structural mechanical behavior. Next, the existing limitations of the Widyowijatnoko's classification are highlighted and examined. Finally, we discuss the relevance of completing the current classification method to guide the decision-making process from the early stages of the design project.

## 2 Design of bamboo connections considerations

The primary function of a connection is its mechanical capacity to transfer loads by bringing together several materials with different properties. However, the design of connections must take into account a number of factors that can interact with each other. In this part, we focus on the key factors of connections design by describing the bamboo properties and underlining the mechanical requirements ensuring their performance. Then, we propose a primary representation of the multidimensional design factors of the bamboo connections.

### 2.1 Material anatomy and hygroscopic properties

**Cross section**
The bamboo culm cross-section is typically defined by a hollow cylindrical form, although some species have solid cross-sections, commonly characterized by their diameter and their wall thickness [2]. However, bamboo, as a vegetal fiber material, presents significant interspecies, intraspecies, and intraspecimen variability in its geometric properties (see Fig. 1 and Table 1). The conical shape of the bamboo culm results in variations in both diameter and wall thickness along its longitudinal axis, with these parameters being larger near the base of the culm [2]. Moreover, the bamboo cross-section is not strictly circular, which can create difficulties in the design, the fabrication, and the standardization of connections.

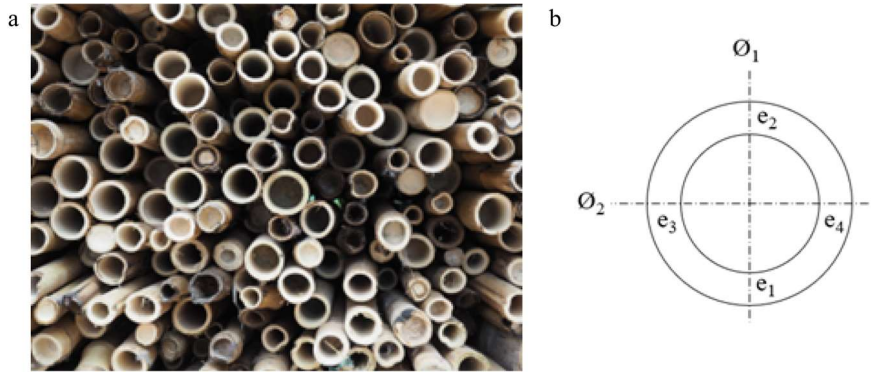

**Fig. 1.** a) Photography illustrating the high variability in bamboo geometry (Photography by the author). b) Diagram based on an on-site survey conducted by the author as part of a characterization project for the species *Guadua angustifolia Kunth* from the Santander region, Colombia.

**Table 1.** Measurements taken from different parts of the culm highlighting the asymmetry of the cross-section as well as the variation in diameter and wall thickness along the culm.

|  | $e_1$ [mm] | $e_2$ [mm] | $e_3$ [mm] | $e_4$ [mm] | $\varnothing_1$ [mm] | $\varnothing_2$ [mm] |
|---|---|---|---|---|---|---|
| Base | 21.0 | 22.0 | 20.0 | 24.0 | 133.0 | 130.0 |
| Bottom section | 16.0 | 17.0 | 17.0 | 15.0 | 128.0 | 121.0 |
| Upper section | 13.0 | 12.0 | 12.0 | 13.0 | 110.5 | 100.5 |

**Nodes**

Another anatomical characteristic of bamboo is the presence of nodes along the culm. These nodes are considered as diaphragms distributed along the length of the culm (see Fig. 2). The average distance between nodes varies depending on species or specimen [2,9]. Generally, the nodes are more tightly spaced at the lower part of the culm and become increasingly spaced as one moves up the length of the culm.

The node's location can be significant in the design and the fabrication of connections. At this anatomical singularity, the fibers are interwoven, providing resistance to transverse tension and shear [9,12]. In South America, some builders also take advantage of these internal diaphragms by injecting mortar into the internodes to create connections [1,9,13].

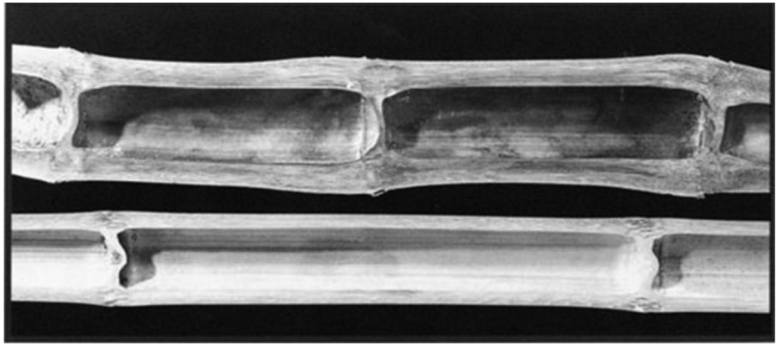

**Fig. 2.** Split bamboo with internodal culm walls surrounding the lacuna and partitioning nodes [14].

**Shrinkage**

Bamboo, as a natural fiber material, is partially composed of cellulose [2], a hydrophilic compound whose properties vary with its moisture content, resulting in anisotropic volumetric shrinkage during drying. The anisotropic shrinkage brings transverse contraction of the bamboo culm, thus generating tensile stresses that can lead to cracking of the culm. Both the transverse contraction and the capacity of the bamboo culm's cross section to vary over time due to shrinkage when the bamboo used in construction has not been properly dried (see Fig. 3), are parameters particularly critical for the design of connections.

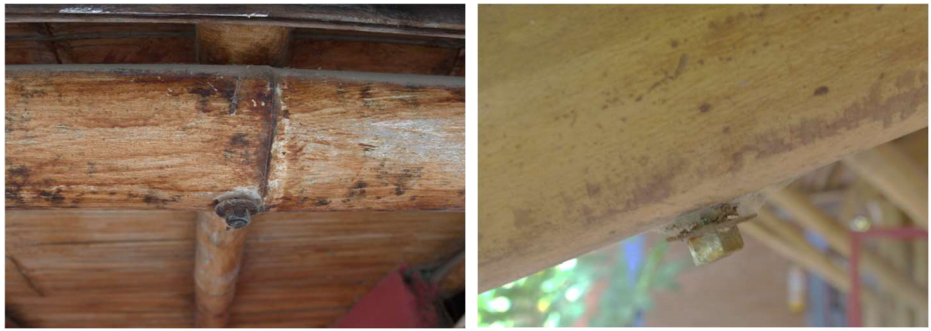

**Fig. 3.** Highlighting the phenomenon of volumetric shrinkage of bamboo. Photographs of structures over 20 years old show that the washer detaches from the outer wall of the bamboo culm, indicating a possible phenomenon of transverse shrinkage and/or settlement (Photography by the author).

## 2.2 Material mechanical properties

**Anisotropy**

The fiber distribution within the bamboo culm makes it a highly anisotropic material and affects its mechanical properties [15]. The fibers are oriented longitudinally in the internodes, giving the material excellent mechanical properties in axial tension [9]. However, the absence of transversely oriented fibers in the internodes makes bamboo particularly weak in longitudinal shear and transverse tension, rendering it highly prone to cracking, which is considered a critical consideration in connection design.

At the nodes, the fibers change direction and intersect (Fig. 4a). This configuration enhances resistance to shear and transverse tension. However, the fiber bifurcation creates a weak point when the bamboo is subjected to axial tension [9]. Nodes are particularly important in connection design as they provide resistance to transverse forces that may arise at these points in the structure. The lack of resistance to transverse forces can lead to the appearance of pathologies and failure in the stressed areas of the connections (Fig. 4b).

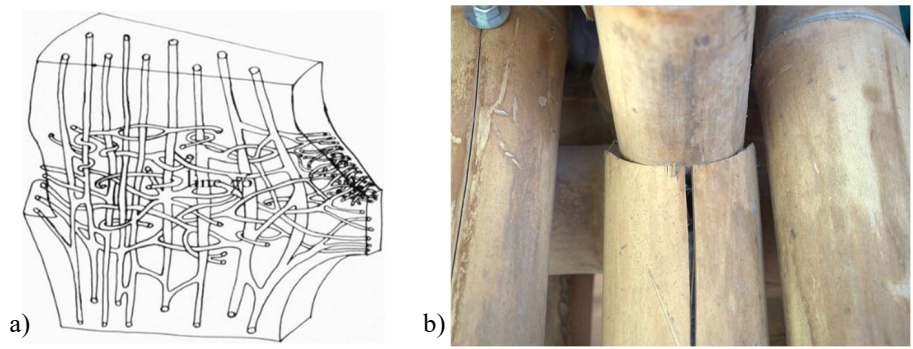

**Fig. 4.** a) Structure of a node with vascular anastomoses [16]. b) Fish mouth joint without a node nearby. Appearance of a crack due to the transverse tensile component generated by the connection (Photography by the author).

**Density**

The fiber density in the bamboo culm wall increases from the inner to the outer part of the culm (Fig. 5). The inner layers are primarily composed of parenchyma cells, which are the weakest tissues in bamboo [2,17]. This consideration will affect the strength of connections relying on force transmission through the interior of the bamboo culm, as its resistance will be limited by the shear strength of the parenchyma layers [18]. Given the low density of the inner layers, when applying transverse compressive forces to the bamboo culm wall for connection purposes, there is a risk of compression of the parenchymatic tissues over time. This phenomenon should be considered in the design of the connection to account for potential long-term effects.

66

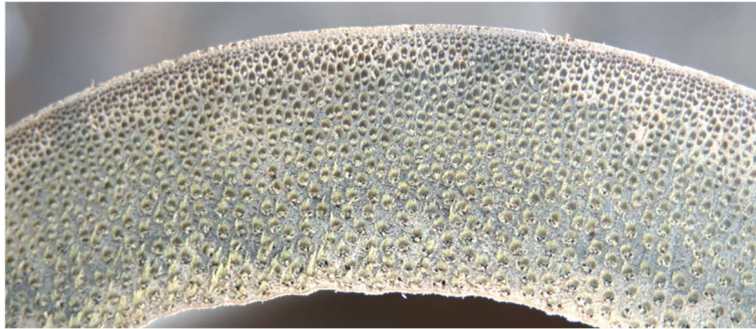

**Fig. 5.** Cross section of a bamboo culm wall showing an increase in fiber density from the inner to the outer part (Photography by the author).

**Failure mode**
Bamboo is a material that exhibits brittle failure [15]. A well-designed structural system can enhance the resilience of such constructions. Similarly, achieving ductile behavior through the design of connections can help prevent brittle failure of the structural components. In seismic regions, the ductile behavior of the connections allows for the dissipation of some of the kinetic energy, thus reducing the forces within the structure.

**Elastoplastic behavior**
Bamboo is an elastoplastic material and is therefore susceptible to creep [17].This phenomenon will induce long-term plastic deformation in structural elements subjected to bending. In the case of connections, creep might affect the long-term stiffness of the joints, which may, in turn, impact the global deformations of the structure as well as the distribution of forces within it. The authors outline a lack of research on this topic, which could be important to consider in the case of large-scale permanent structures.

### 2.3　Mechanical requirements for connections

Connections are essential for transferring forces between the elements composing a structure. Joints used in bamboo structures facilitate the transfer of compressive, tensile, and shear forces. However, similar to timber connections, the transfer of bending moments is challenging to achieve with bamboo [11,19]. Thus, it is preferable to complete the global stability of structures through triangulation, creating stable frames within their plane, rather than relying on the design of moment-resisting connections. Additionally, bamboo exhibits very limited torsional resistance, as torsion induces longitudinal shear stresses that designers must avoid. Here we underline the relevant mechanical considerations to integrate in the design of the bamboo connections.

**Load type and load capacity**
Before evaluating the load capacity of the connection, it is important to properly identify the types of forces it will be subjected to: axial compression, transverse compression, axial tension, transverse tension, and shear. Due to the anisotropy of bamboo, it is particularly sensitive to transverse forces, which can significantly reduce the connection's load capacity and result in brittle failure. This will be a key consideration in the design process. The load capacity of the connection is a fundamental parameter in the design of the structure, as it forms the basis for evaluating the connection's resistance to the applied forces.

**Axial stiffness and initial slippage**
One critical property of bamboo connections is their axial stiffness, which significantly influences the evaluation of the global deformations of the structure. Thus, for mechanically assembled elements, such as beams composed of multiple culms or laminated bamboo strips, the connection stiffness directly impacts the overall rigidity of the composite element [20,21]. Additionally, it is important to note that the connection stiffness may evolve over time due to long-term phenomena such as creep.

Another important aspect to consider is the initial fitting gap in certain types of connections. For bolted connections, the Colombian standard recommends drilling with a tolerance of 1.5 mm relative to the bolt diameter [4]. This gap, necessary for assembly, directly affects the global deformations of the structure and the distribution of forces among its components, potentially leading to stress concentrations at specific points in the structure.

**Cracking risk induced by the assembly**
Due to its low shear and transverse tensile strength, bamboo is a highly crack-sensitive material. In specific cases, the cracking can significantly impact the load-bearing capacity of structural elements. As specified in ISO 22156 §10.7, the risk of the assembly causing cracking must be taken into account during the design process. For instance, bolted connections generate stress concentrations and transverse tensile stresses in the bamboo wall, increasing the risk of long-term cracking. Additionally, if the force transfer through the bolt includes a transverse component, the risk of cracking is further heightened (see Fig. 6).

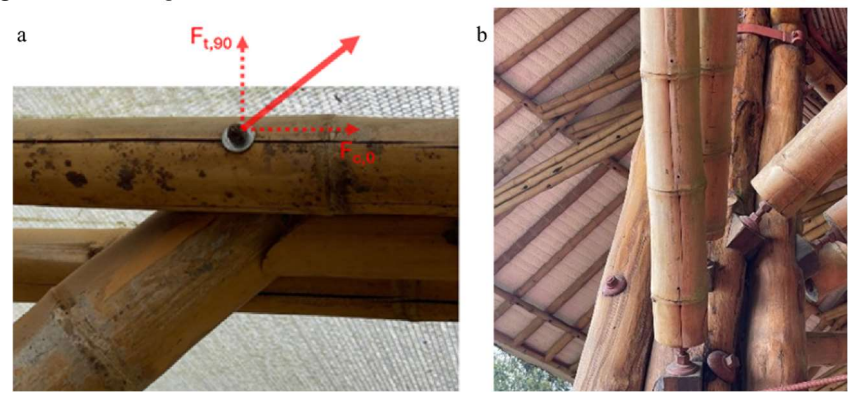

**Fig. 6.** a) Transverse forces ($F_{t,90}$) in the connection generate transverse tensile stresses within the horizontal culms, leading to splitting and a reduction in the stiffness and strength of the connection. b) Long-term crack formation caused by the simple perforation of the culm and hygroscopic shrinkage (Photographs by the author).

Connections transferring load through the internal wall of the bamboo culm generates transverse tensile stresses that can lead to crack formation when shear forces are involved. Another common issue with this type of connection is the development of cracks caused by the restriction of hygroscopic shrinkage due to the insertion of a rigid element into the internode of the culm [13] (see Fig. 7).

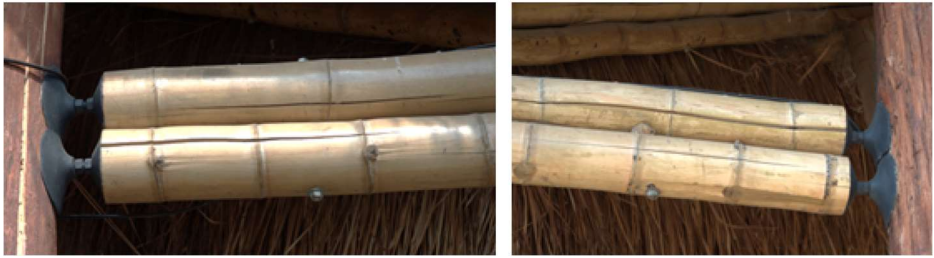

**Fig. 7.** Cracking caused by a combination of the shear component and the restraint of hygroscopic shrinkage related to the injection of mortar (Photographs by the author).

**Ductility**
Bamboo is a brittle material, making the pursuit of ductility in connections critical to achieving a ductile failure mode and ensuring the safety of designed structures. In seismic regions, enhancing ductility through connections facilitates the dissipation of kinetic energy and participating in the reduction of internal forces within the structure. More broadly, achieving a ductile behavior also allows for adjustments to safety factors, which can significantly impact the overall cost-efficiency of the project. ISO 22156 provides recommendations on safety factors to be considered based on the ductility level of the connections. Several studies have highlighted the brittle behavior of bamboo connections [13,18]. In the case of bolted connections, a recent study demonstrated that the use of metal collars can provide a ductile behavior for this type of connection while also offering some resistance to transverse tension [10].

### 2.4 Building methods

**Dependency on local know-how**
In some regions of Asia or Latin America, bamboo craftsmanship is an integral part of the construction heritage [1]. The presence of highly skilled labor makes it possible to integrate traditional joinery, such as fish-mouth joints, into architectural design. In regions where this expertise is not locally available, simpler connection methods are necessary. This is one of the reasons why bolted connections are increasingly used, despite the various structural challenges they may pose. Considering the locally available craftsmanship will guide the designer's choices in relation to feasibility criteria and quality execution required.

**Assembly**
The construction method of the structure also impacts the design of the connections. Traditional connections, such as fish-mouth joint, commonly used in Latin America, make it difficult to prefabricate the structural elements. Each connection must perfectly fit the unique geometries of the bamboo pieces to be joined, making workshop standardization of fabrication process difficult, if not impossible. Threaded rod connections, on the other hand, offer the advantage of adapting more easily to the irregular geometry of bamboo and require less skilled labor. They also allow multiple culms to be joined together to create composite sections [9], although the composite effect induced by the connection is to be considered as nonexistent for the structural design [5].

Another aspect to consider is the disassembly capacity of structures, which can become an overriding criterion in the design of the connection. Disassembly can occur in the case of temporary structures, for the reuse of materials at the end of the building's life, or to facilitate the maintenance and the replacement operations, increasing the resilience of the structure against attack by fungi or insects such as the bamboo borer *(Dinoderus minutus)*.

**Calculation method**

The ISO 22156 standard provides guidelines for the design methods of bamboo connections. It includes calculation formulas for contact-based connections as well as bolt connections [5]. These formulas, although conservative, allow for the analytical assessment of connection strength without the need for mechanical testing, which can sometimes incur significant costs. Thus, for other types of connections, mechanical testing is required, which can be conducted either by the complete joint testing method or by the component capacities method [5].

The Colombian standard NSR-10 provides design rules for both traditional and bolted connections. For bolted connections, it includes tables with allowable strength values for different load directions: P (axial), Q (transverse), and T (force perpendicular to the fibers of one connected element and parallel to the fibers of the other). However, the values for transverse loading (Q) raise concerns due to the assumed low tensile strength of bamboo in transverse tension and the risk of long-term cracking. There is also a lack of integration of the diversity of the bamboo resource in the Colombian standard, which is limited to the *Guadua angustifolia Kunth* species.

## 2.5 Integrative design approach

The previous sections have highlighted the importance of considering the geometric, anatomical, and hygroscopic properties of bamboo, as well as the mechanical requirements of the structure design and the fabrication specifications, when designing connections. To these initial design factors, we can add architectural expression criteria, if the connections are visible or not, environmental criteria, assessed using life cycle analysis methods or by material consumption calculation, or cost criteria. The multidisciplinary approach of the design process, from its early stages, refers to the concept of integrated design, which has been introduced to the AEC industry to enhance the cost efficiency of projects as well as, today, enhance the sustainability of the design [22]. The literature review clearly shows that such approaches should be favored for the design of bamboo structure connections, as the response to the transfer of loads cannot be considered as the only design criterion, particularly given the impact of the material properties and their dependence on the quality of the building methods.

The authors summarize, in the diagram presented in Fig. 8, the key technical factors to consider in the design of bamboo connections, identified in the literature review. This representation can be completed in future work by integrating objective criteria and specifying the interdependent relationships existing between the different factors highlighted, which may lead to new guidelines for designers.

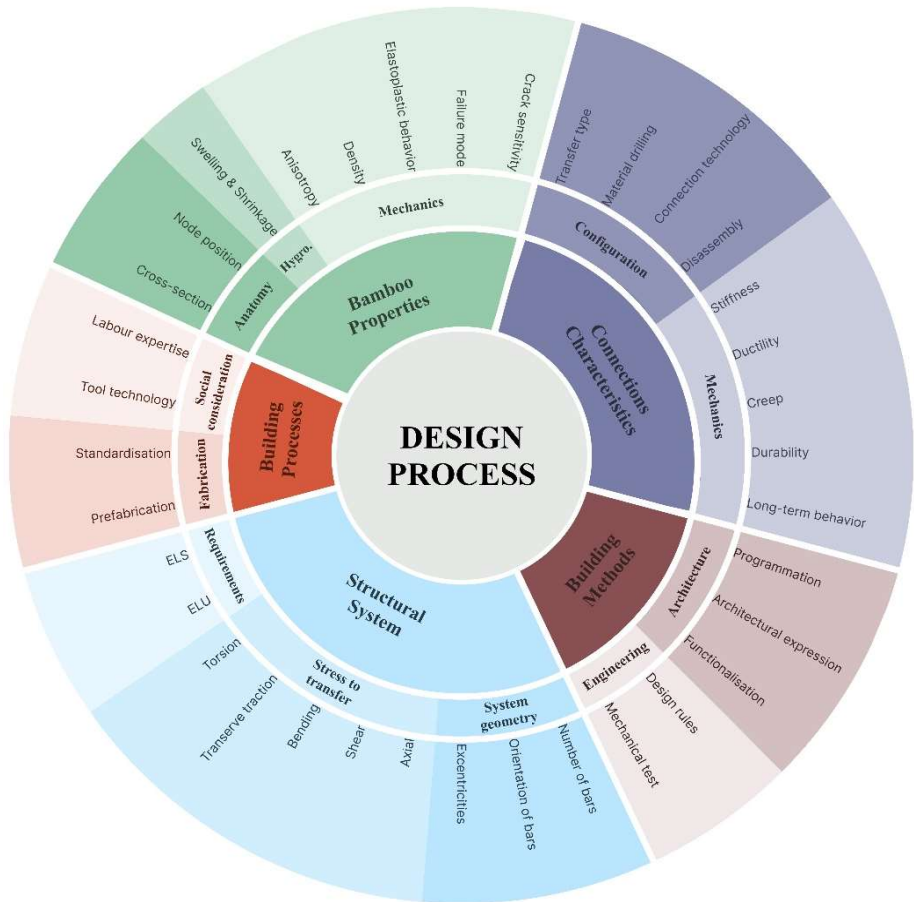

**Fig. 8.** Overview diagram of the technical considerations to integrate in the design process of bamboo connections, proposed by the author.

## 3      Connection classifications

The study of bamboo connections remains relatively underrepresented in the scientific literature, despite increasing research efforts [6-11]. Connections are often classified into two main categories: traditional and modern [9]. The traditional connections refer to the one used in traditional and vernacular craftsmanship knowledge, whereas the modern connections relate to engineering products. However, this classification offers no insight into the mechanical behavior of the connections. Janssen [12] proposed a classification system based on load transfer mechanisms, later refined by Widyowijatnoko [9]. The refined classification system defines six load transfer mechanisms that can be combined to form a structural connection (see Fig. 9).

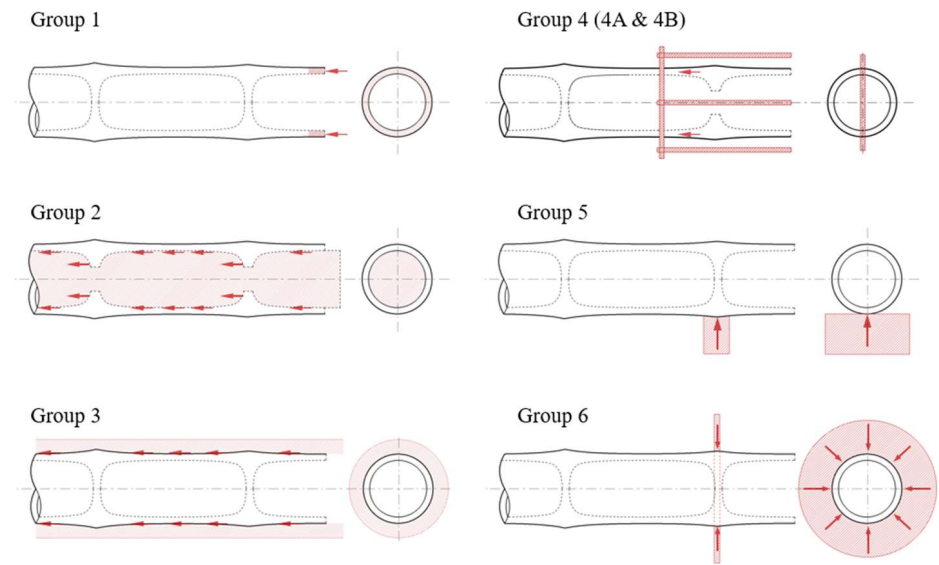

**Fig. 9.** Classification of connections according to Widyowijatnoko. (Adapted from [9]). Group 1: Transferring compression through contact with the whole section. Group 2: Transferring force through friction on the inner surface or compression to the diaphragm. Group 3: Transferring force through friction on the outer surface. Group 4: Transferring force through bearing stress and shear to the bamboo wall from perpendicular elements connected from the inside (4A) or outside (4B) of the culm. Group 5: Transferring force perpendicular to the fibers. Group 6: Transferring radial compression to the center of the culm through shear and circumferential stress perpendicular to the fibers.

## 4      Analysis of the of the existing classification limitations

The classification proposed by Widyowijatnoko provides a clear understanding of the load transfer mechanisms within connections and assumes that most connections result from a combination of multiple load transfer types. While this classification is insightful, it does not complete the criteria to integrate in the connection design process, which interact with the mechanical behavior of the connections and supplement it with other decision-making criteria. The authors underline and discuss four key factors, particularly associated with the quality of execution as well as the material anatomy and properties, which affect the mechanical behavior of the connections.

### 4.1 The role of the technical expertise and the complexities of implementation

Group 1 transfers loads through contact with the bamboo culm's cross section, being most effective for axial compression when the cut is perpendicular to the fibers. The fish-mouth joint, commonly used in the traditional and vernacular bamboo architecture, is highly sensitive to the execution quality. Its semi-circular shape creates transverse tensile stresses, potentially causing culm cracking (see Fig. 10).

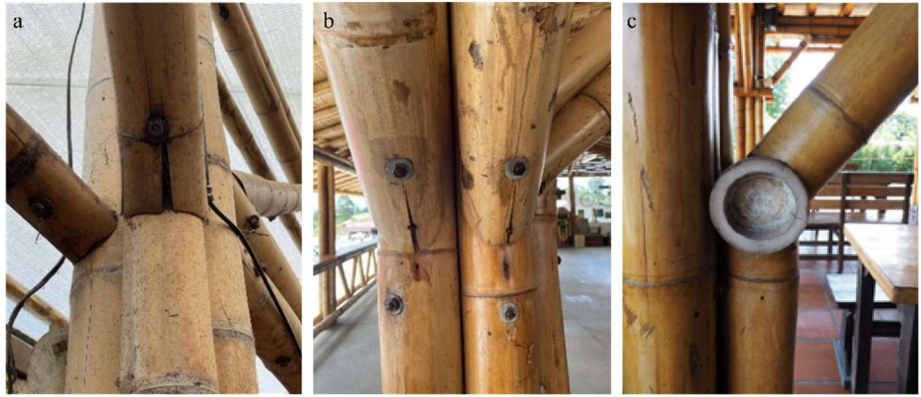

**Fig. 10.** a) Crack caused by transverse tensile stresses resulting from improper joint execution. b) Crack caused by the insertion of nails and the absence of a node near the joint area. c) Properly executed fish-mouth connection (Photographs by the author).

In contrast, group 4 used a connector placed perpendicularly to the longitudinal axis of the bamboo culm to transfer load. This increasingly used method is similar to bolt-type connections in timber construction and offers the advantage of being quick to implement without the need for specialized labor. It also adapts to the geometry of the culm, facilitating the standardization of joints. Furthermore, recent research has shown that, based on Johansen's theory [23], reliable analytical formulas can be derived to calculate the strength of the bold-type joints with mortar injection [8].

### 4.2 The culm cross-section variability

Group 2 involves load transfer through the internal wall of the bamboo culm, allowing for a discrete joint that can function in tension, compression, and shear. However, this mechanism presents several challenges. The first is adapting to the internal geometry of the culm which can vary significantly. Another issue is the adhesion to the internal wall, which consists mainly of parenchyma and has low resistance [17]. Inserting a solid element inside the culm can restrict hygroscopic shrinkage, creating transverse tensile stresses in the bamboo wall, leading to cracks. Additionally, when shear forces are involved, the equilibrium of the joint generates transverse stresses that can also cause cracking.

Group 3 transfers loads through the external wall of the bamboo culm. This method typically helps reduce crack formation by applying perimeter pressure on the wall.

However, since the external wall contains silica, friction forces are limited [9], reducing the joint's resistance and axial stiffness. Additionally, the non-circular geometry of the culm can introduce bending stresses in the wall, affecting the joint's overall strength. In the case of rigid metal collars, stress concentrations may arise at the contact point with the culm's external wall.

### 4.3 The node position in the connection design

Group 5 involves load transfer perpendicular to the fibers. Applying a perpendicular compression generates bending stresses, leading to transverse tensile stresses in the culm wall, which can cause brittle failure by crushing the bamboo (see Fig. 11.a). This method has limited load-bearing capacity and is prone to brittle failure. Typically, such joints are made near a node, where the diaphragm provides the culm with some resistance to transverse compression (see Fig. 11.b).

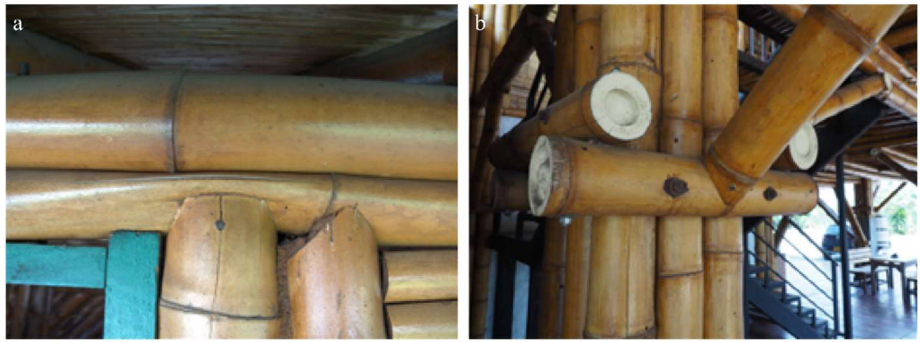

**Fig. 11.** a) Bamboo collapse under transverse compression load. b) Brace of a portal-type structure resting on a node to avoid crushing (Photographs by the author).

Another method involves inserting a rigid element at the internode, such as mortar or wood. However, since the contact between the internal element and the bamboo wall is not perfect, and the internal layers of the bamboo wall are low-density and may compact, long-term settling may occur. While this does not affect the joint's strength, it may impact the distribution of forces within the structure and influence global deformations over time.

Finally, due to the unique geometry of the bamboo culm, the contact area is often very small, and significant stress concentrations can occur at the contact point between the two connected elements, increasing the risk of crushing the bamboo fibers.

### 4.4 The lack of consideration for cracks and long-term behavior

Group 4, unlike the other one, require perforating the bamboo culm wall, making this assembly method particularly prone to cracking. Cracks can arise from various factors, including transverse tensile stresses, hygroscopic shrinking of the culm, and stress concentrations. Best practices include placing a node between the connector and the culm end or avoiding aligning multiple connectors in a straight line. Various reinforcement methods can provide transverse resistance at the connection points, thus reducing the risk of cracking in the bamboo culm. Arce-Villalobos presents a connection using a glued wood insert inside the internode of the bamboo culm and notes that the adhesive bond provides some transverse resistance at the connection [18]. However, this resistance remains limited due to the low strength of the inner layers of the culm. Another commonly used method involves placing a metal strap around the area subjected to transverse tension (see Fig. 12.b). However, the long-term effectiveness of this method can be questioned, as hygroscopic shrinkage may affect the bond between the collar and the bamboo culm. Other studies have focused on transverse reinforcement through the addition of fiberglass (see Fig. 12.c) [24].

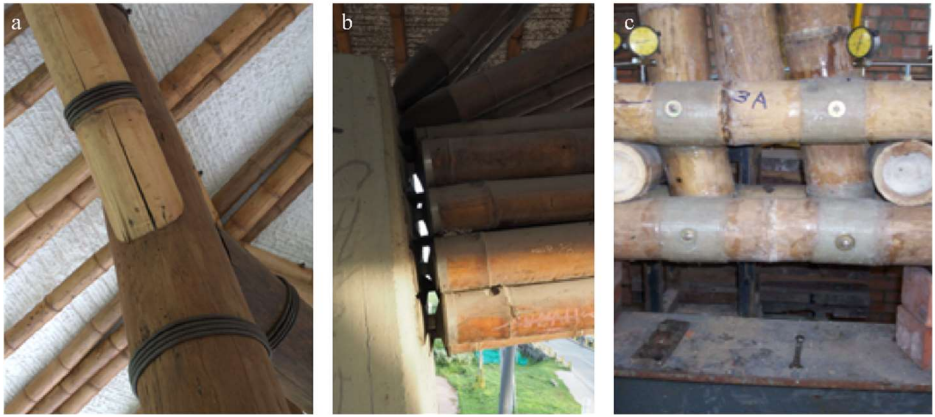

**Fig. 12.** a) Transverse reinforcement using welded rebar (Photography by the author). b) Transverse reinforcement using stainless steel bands (Photography by the author). c) Transverse reinforcement using fiberglass [24].

The hygroscopic shrinkage also affects Group 3, especially if the bamboo is inadequately dried. Over time, this shrinkage can lead to a loss of adhesion between the joint element and the culm's external wall, compromising the joint's resistance and stiffness. Group 6, based on load transfer through radial compression toward the center of the bamboo culm, enables the transmission of tensile forces. Widyowijatnoko designed the only one specific connection type in this group, increasing the axial tensile efficiency [9]. He describes the failure mode as partially ductile. The long-term behavior of this connection typology raises questions, particularly concerning axial stiffness. Changes in the culm geometries due to the hygroscopic shrinkage could lead to a reduction in axial stiffness over time, potentially resulting in excessive global deformations within the structure.

# 5 General discussion

## 5.1 The need for extension of the Widyowijatnoko classification

The classification of connections proposed by Widyowijatnoko [9] provides a clear understanding of the different load transfer mechanisms existing in bamboo connections.

However, when this classification is examined in light of the technical considerations described in Section 2, it becomes clear that it fails to capture the complexity of connection design and its inherently multi-criteria nature. For instance, connections in Group 3 raise significant concerns about their long-term reliability and effectiveness, particularly when considering the risk of hygroscopic shrinkage and creep.

Finally, given the specificities of the material, it seems important to enhance this classification by incorporating the relationship with the intrinsic properties of the material and linking it to local techniques and know-how.

It therefore seems necessary to expand the current classification of connections in order to holistically guide the design process and thus ensure the mechanical performance and durability of bamboo structures.

## 5.2 The need for strengthen the material and connection properties link

This paper highlights the specific properties of bamboo as a construction material and their impact on the design of connections. Its high anisotropy makes it highly susceptible to cracking, a phenomenon that can be exacerbated by the occurrence of transverse stresses in the connections, an aspect that has not been extensively studied in the literature. This raises questions about the suitability of bolted connections for the construction of large-span permanent structures.

Hygroscopic shrinkage and swelling of the material must also be taken into account during the design process of the connections. This issue is particularly problematic for connections in Groups 2 and 3, as it can significantly reduce load-bearing capacity and, in some cases, lead to brittle failure of the connection.

Finally, long-term effects, such as creep, could also impact the stiffness of the connections, altering the distribution of forces within the structure. Currently, there is limited research aimed at assessing the long-term behavior of bamboo connections, a crucial area of study to ensure the viability of structures over time.

These considerations highlight the importance of the relationship between the material properties and those of the connection, which must be carefully considered in the design of connections for bamboo structures.

### 5.3 The need to confront the design with the practical implementation

The economic and cultural context of the project will also have a significant influence on the design of connections.

The availability of certain construction materials or tools affects the feasibility of specific connection types. Traditional connections typically require hand-tool-technology (see Fig. 13.a), whereas more modern connections may involve machine-tools-technology (see Fig. 13.b) or event information-tool-technology, which may not always be available in certain regions of the world.

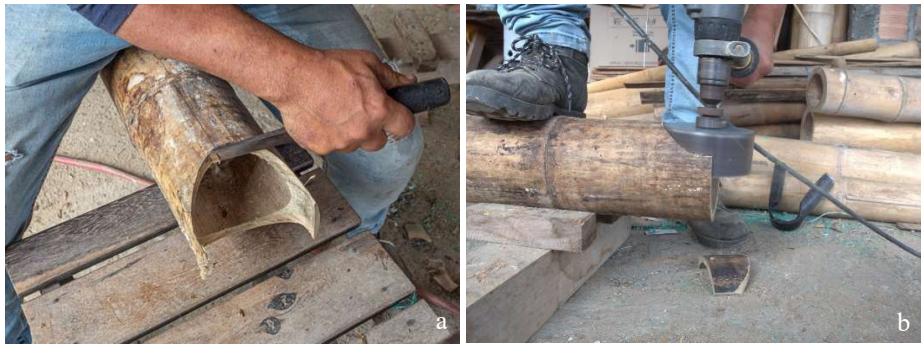

**Fig. 13.** a) Hand-tool-technology. b) Machine-tool-technology. (Photographs by the author).

Similarly, the presence of local craftmanship plays a key role. Some connection types require specialized knowledge tied to local construction practices, making them difficult to transpose to other regions. These local skills are influenced by numerous socio-cultural factors, as well as by the specific properties of endemic bamboo species. The coffee region of Colombia, for instance, possesses unique ancestral craftsmanship, adapted to the use of the local species, *Guadua Angustifolia Kunth*.

The traditional connection requires a high level of craftsmanship, which is essential to consider in the design of structures. Skilled craftsman typically possesses extensive knowledge of the material and its specific characteristics. Co-designing the connections between the engineer, architect, and craftsman thus becomes a key concept in the development of the project. The work of Colombian architect Simon Vélez is a perfect example of this approach.

In contrast, bolted connections require little craftsmanship for their implementation. The engineer's role becomes crucial in the design of the structure. The design must specifically aim to prevent the occurrence of transverse stresses in this type of connection. The presence of clearance in the connection must also be considered in the analysis of the structure's overall displacement.

In general, due to the material's specific properties and its close relationship with available local craftsmanship, it is essential for the designer to engage with the site conditions to fully grasp the nuances of connection implementation and their overall impact on structural efficiency.

## 6   Conclusion

This research proposes a review of the material, the mechanical, the building methods and processes considerations involved in the design of the connections of bamboo structures. The aim is to provide guidelines to designers to support the broader adoption of bamboo in construction and ensure the durability and sustainability of structures.

The authors have proposed an overview diagram which underline the technical considerations involved in the connection design process, based on the literature review and the practical expertise. The review highlighted that the connection design process is multi-criteria and strongly impacted by the intrinsic bamboo properties and the quality achievable in execution.

Furthermore, the analysis of the Widyowikatnoko's classification, currently used as guidelines by designers, emphasizes that, although relevant, it overlooks criteria that nevertheless impact on the design outcomes, including structural efficiency and durability, that can lead to the onset of pathologies. This research clarifies a need to complement these classifications in order to support designers in their work to achieve sustainable, high-performance connections by the integration of other decision factors than the force transfer alone.

Further research will focus on extending the Widyowikatnoko's classification, with particular attention to the interrelationship between the material properties of bamboo, the local tools, techniques and know-how, and then the mechanical efficiency of the bamboo structure connections for which work remains to be carried out.

## References

1. Hidalgo-Lopez, O. The Gift of the Gods. D'Vinni Ltda, Bogota, Colombia (2003).
2. Liese, W., Köhl, M. Bamboo. The Plant and its Use, Tropical Forestry. Springer Cham (2015).
3. Kaminski, S., Lopez, L.F., Trujillo, D., Zea Escamilla, E., Correa-Giraldo, V., Correal Daza, J. Composite Bamboo Shear Walls – a Shear Wall System for Affordable And Sustainable Housing in Tropical Developing Countries. In : SECED 2023 Conference Proceedings, Cambridge, UK (2023).
4. AIS NSR – 10. Reglamento Colombiano De Construccion Sismo Resistente, Bogota, Colombia (2010).
5. ISO 22156 : 2021 Bamboo structures – Bamboo culms – Structural Design (2021).
6. Trujillo, D. Axially loaded connections in Guadua bamboo. In : 11th International Conference on Non-conventional Materials and Technologies : NOCMAT 2009, University of Bath, UK (2009).
7. Malkowska, D., Trujillo, D., Toumpanaki, E., Norman, J. Study of screwed bamboo connection loaded parallel to fibre. Construction and Building Materials, 398 : 132532 (2023).
8. Correal, J.F., Prada, E., Suarez, A., Moreno, D. Bearing capacity of bolted-mortar infill connections in bamboo and yield model formulation. Construction and Building Materials, 305 : 124597 (2021).
9. Widyowikatnoko, A. Traditional and innovative joints in bamboo construction. PhD Thesis, Faculty of Architecture of the RWTH Aachen University, Germany (2012).

10. Paraskeva, T., Pradhan, N.P., Stoura, C.D., Dimitrakopoulos, E.G. Monotonic loading testing and characterization of new multi-full-culm bamboo to steel connections. Construction and Building Materials, 201 : 473 – 483 (2019).
11. Harries, K.A., Rogers, C., Brancaccio, M. Bamboo joint capacity determined by ISO 22156 'complete joint test' provisions. Advances in Bamboo Science, 1 : 100003 (2022).
12. Janssen, J.J. Designing and building with Bamboo. In : Technical report n°20, International Network for Bamboo and Rattan (2000).
13. Forero, E. Uniones a tension en guadua con mortero y varilla. Comportamiento de uniones con uso de expansivo en el mortero. Master Thesis, Universidad Nacional de Colombia, Bogota, Colombia (2003).
14. Liese, W. The anatomy of bamboo culms. In : Technical report n°18. International Network for Bamboo and Rattan (1998).
15. Kaminski, S., Lawrence, A., Trujillo, D., Felthman, I., Lopez, L.F. Structural use of bamboo. Part 3 : Design values. The Structural Engineer, 94(12) : 42-45 (2016).
16. Ding YuLong, D.Y., Liese, W. Anatomical investigations on the nodes of bamboos. In : The Bamboos. Proceedings of an International Symposium, pp. 269-283. Academic Press for the Linnean Society of London, UK (1997).
17. Janssen, J.J. Bamboo in building structures. PhD Thesis, TU/e Eindhoven University of Technology, Netherlands (1981).
18. Arce-Villalobos, O.A. Fundamentals of the design of bamboo structures. PhD Thesis, TU/e Eindhoven University of Technology, Netherlands (1993).
19. Camacho, V.P.I. Estudio de connexiones en guada sollicitadas a momento flector. Master Thesis, Universidad Nacional de Colombia, Bogota (2002).
20. NF EN 1995-1-1 : Eurocode 5 – Conception et calcul des structures en bois – Partie 1.1 : Généralités – Règles communes et règles pour les bâtiments. AFNOR éditions (2005).
21. Correal, J.F., Calvo, A. Flexural behavior of composite section of multiple full-culm bamboo beams with mortar infill bolted joints. Construction and Building Materials, 412 : 134785 (2024).
22. Knippers, J., Kropp, A., Menges, A., Sawodny, O., Weiskopf, D. Integrative computational design and construction : Rethinking architecture digitally. Civil Engineering Design, 3(4) : 123-134 (2021).
23. Johansen, K.W. Theory of timber connections. International Association of Bridge and Structural Engineering Publication, 9 : 249-262 (1949).
24. Echeverri Gómez, A. Comportamiento de unions en Guadua utilizando fibra de vidrio. Master Thesis, Universidad Nacional de Colombia, Bogota (2007).